\documentclass[12pt,preprint]{aastex}

\usepackage{emulateapj5}
\usepackage{onecolfloat}
\usepackage{graphicx}
\usepackage{times}

\def\gsim { \lower .75ex \hbox{$\sim$} \llap{\raise .27ex \hbox{$>$}} }
\def\lsim { \lower .75ex \hbox{$\sim$} \llap{\raise .27ex \hbox{$<$}} }

\shorttitle{Velocity trends in the debris of Sagittarius and the shape
of the dark-matter halo of the Galaxy} 
\shortauthors{A. Helmi}

\begin{document}

\twocolumn[

\title{Velocity trends in the debris of Sagittarius and the shape
of the dark-matter halo of the Galaxy}

\author{Amina Helmi\altaffilmark{1}}
\affil{Kapteyn Institute, P.O. Box 800, 9700 AV Groningen, The Netherlands}

\begin{abstract}
 Recently, radial velocities have been measured for a large sample of
M giants from the 2MASS catalog, selected to be part of the Sgr dwarf
leading and trailing streams. Here we present a comparison of their
kinematics to models of the Sgr dwarf debris orbiting Galactic
potentials, with halo components of varying degrees of flattening and
elongation.  This comparison shows that the portion of the trailing
stream mapped so far is dynamically young and hence does not provide
very stringent constraints on the shape of the Galactic dark-matter
halo. The leading stream, however, contains slightly older debris, and
its kinematics provide for the first time direct evidence that the
dark-matter halo of the Galaxy has a prolate shape with an average
density axis ratio within the orbit of Sgr close to 5/3.
\end{abstract}

]
\altaffiltext{1}{E-mail: ahelmi@astro.rug.nl}

\section{Introduction}
\label{sec:intro}

Using M giant candidates from the 2MASS catalog, Majewski et
al. (2003) have been able to trace streams from the Sagittarius dwarf
through roughly 360 degrees on a great circle on the sky. More
recently, the radial velocities of a subset of several hundred of
these M giant candidates have been measured. The results have been
presented in Law et al. (2003) and in tabular form (for the trailing
stream) in Majewski et al. 2004 (hereafter M04).  Hence 4 of the 6
phase-space coordinates of the stars in the streams are now available
for a comprehensive dynamical study.

One of the intriguing questions that can in principle be addressed
with such a dataset, concerns the shape of the dark-matter halo of the
Galaxy. Is this flattened? Prolate or oblate? Ibata et al. (2001) and
Majewski et al. (2003) have argued that the fact that the debris of
Sgr is distributed along an almost complete great circle on the sky
implies that the dark halo of the Milky Way has to be very close to
spherical, with allowed (density) axis ratios in the range 0.9 to 1
(Law et al. 2003). On the other hand, Helmi (2004) has shown that, if
the debris discovered so far is made up of material lost in the last 3
to 4 pericentric passages (i.e. less than 3 Gyr ago), then such debris
would also be found along a great circle on the sky. Flattened
(oblate) or elongated (prolate) halos with average density axis
ratios, measured within the region probed by the debris, that are as
low as 3:5 or as high as 5:3, respectively, would be allowed. Such
values would be consistent with those found in numerical simulations
of cold dark-matter halos (e.g. Bullock 2002).

The motivation behind this Letter is to establish whether the
additional kinematic information that has now become available could
perhaps break this degeneracy and allow us to better constrain the
axis ratio of the Galaxy's dark-matter halo.

\section{Comparison of models and data}
\label{sec:arcts}

The models presented here are more thoroughly discussed in Helmi
(2004). Now we only present a brief summary. We have performed
numerical simulations of the disruption of a system like the Sgr dwarf
orbiting a Galactic potential with 3 components: a bulge, a disk and a
dark logarithmic halo

\begin{equation}
\label{sag_eq:halo}
\Phi_{\rm halo} = v^2_{\rm halo} \ln (R^2 + z^2/q^2 + d^2),
\end{equation}

where $d =12$ kpc and $v_{\rm halo}= 131.5$ km/s (Johnston et
al. 1996). The parameter $q$ is allowed to vary from 0.8 to 1.25, that
is, from an oblate to a prolate configuration\footnote{It is worth
noting that the isodensity surfaces are more oblate (prolate) than the
equipotential surfaces.}. The satellite galaxy is modeled by a set of
50000 self-gravitating particles. We use a quadrupole expansion of the
internal potential of the system, and choose a King profile for the
pre-disruption dwarf. The orbital initial conditions are chosen to
satisfy the constraints given by the present position and radial
velocity of the main body of Sgr (Ibata et al. 1997). For each of the
$q$ values of the dark halo potentials, we select orbits which have
similar (mean) pericenter and apocenter distances as well as
comparable $L_z$ ($z$-component of the angular momentum) satisfying
the above mentioned constraints. Our models of the dwarf are also
slightly readjusted for each of the dark halo shapes, so as to produce
the same remnant system by the present day. This implies that
differences in the characteristics of the debris may only be
attributed to a change in the flattening of the potential. The other
possible free variables, such as the model of the dwarf and the
orbital parameters, are (essentially) the same by construction.

We have studied five flattenings for the dark halo potential $q =
[0.8, 0.9, 1.0, 1.11, 1.25]$, which correspond to average density axis
ratios (within the orbit of Sgr) $\langle q_\rho \rangle \sim
[3/5,4/5,1$, $5/4,5/3]$. In this Letter we focus on the properties of
the satellite debris after 10 Gyr of evolution.

\subsection{Young streams}

\begin{figure*}[t]
\begin{center}
\includegraphics[height=0.7\textheight,angle=270]{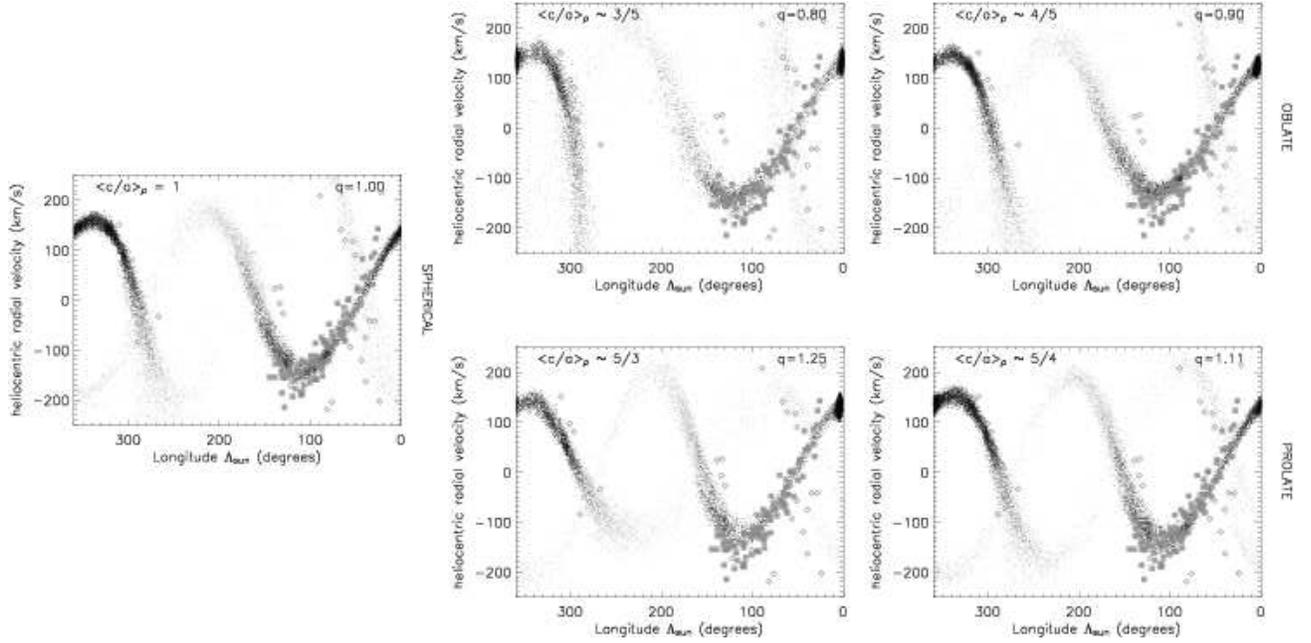}
\end{center}
\caption{Distribution of ($\Lambda_{\rm sun},V_r$) for particles after
10 Gyr of evolution, for each one of the simulations with different
degrees of flattening for the Galactic dark matter halo (black
dots). Particles that have been released in the last 2 Gyr are shown
in black, while the gray dots corresponds to those lost between 2 and
4 Gyr ago. The large gray symbols denote the stars observed by
Majewski et al. (2004) with distances larger than 13 kpc, which is the
criterion used by these authors to minimize the contamination by
foreground thick disk and halo stars.  Solid symbols correspond to
stars which have been considered by M04 as likely members of the
streams, while open symbols denote those that have been statistically
identified as outliers by these authors. The (random) velocity errors
quoted in Majewski et al. 2004 are $\sim$ 6 km/s.
\label{fig:vr}}
\end{figure*}

We study now the characteristics of young debris streams.  We select
those particles that have been released by the dwarf in the last three
pericentric passages (less than 2 Gyr ago in our models). The panels
in Figure~\ref{fig:vr} show the heliocentric radial velocity vs.
longitude $\Lambda_{\rm sun}$\footnote{This variable corresponds to
the angular distance along the stream. Its relation to $(l,b)$ has
been derived in Majewski et al. (2003)} for this set of particles and
for the different shaped dark halos. The large gray symbols in each of
the panels correspond to the stars in Table 3 of M04 with distances is
larger than 13 kpc. There appears to be very good agreement between
the models and the stars in the streams of Sgr observed by M04.


\begin{figure}[t]
\begin{center}
\includegraphics[height=0.27\textheight]{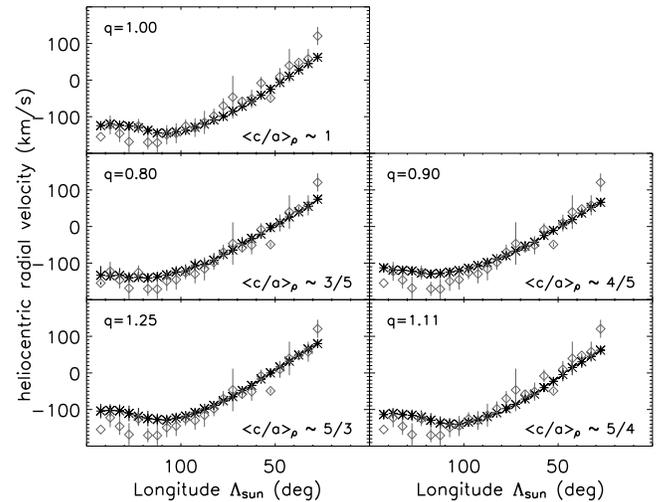}
\end{center}
\caption{The predicted (asterisks) and observed (gray diamonds)
distribution of mean heliocentric radial velocity measured in bins of
$\Delta \Lambda_{\rm sun} = 5^{\rm o}$ width, for the particles (in black)
of Fig.\ref{fig:vr} and for the stars shown as filled gray symbols in
the same figure.
\label{fig:chiq}}
\end{figure}

To quantify how good the agreement is, we perform a $\chi^2$ test. We
bin the particles in the simulations (again only those lost in the
last 3 pericentric passages) in elements of $\Delta \Lambda_{\rm sun}
= 5^{\rm o}$ width, and measure the mean radial velocity and the dispersion
in each bin. We proceed in a similar way for the stars in M04 sample,
selecting those that are more likely members of the stream as derived
by M04 (these are the solid gray points in Figure~\ref{fig:vr}). We
then measure $\chi^2$ as
\begin{equation}
\chi^2 = \sum_{i=1}^{N_{bins}} \frac{(\bar{V}_{i,o} -
\bar{V}_{i,m})^2}{\sigma_{i,o}^2 + \sigma_{i,m}^2}
\label{eq:chiq}
\end{equation}
where $\bar{V}_i$ is the mean radial velocity in the $i$th bin, and
$\sigma_{i}$ its dispersion. The subscripts $o$ and $m$ correspond to
the observational data and to the model points, respectively.  The
results of this test are shown in Figure~\ref{fig:chiq} and in the
last column of Table~\ref{tab:chiq}.

Figure~\ref{fig:chiq} confirms that in all cases, the models provide a
good representation of the data: The value of the reduced $\chi^2$
(which we defined as $\chi^2/N_{bins}$) is close to unity. The most
favored case (taken as that with the smallest $\chi^2$) seems to be
the most flattened halo $q = 0.8$.  A bin to bin analysis of
Figure~\ref{fig:chiq} shows that the bin located at $\Lambda_{\rm sun}
= 52.5^{\rm o}$, which contains only one data point, produces the largest
deviation. If this star is removed, the $\chi^2$ is significantly
reduced for all shapes. In this case, the most favored models are
those with $q = 0.8$ ($\langle c/a \rangle_\rho \sim 3/5$) and $q =
1.25$ ($\langle c/a \rangle_\rho \sim 5/3$). These correspond to the
most oblate and to the most prolate halos we have considered.

We have also measured the $\chi^2$ obtained by comparing models to
models for the same range of longitude $\Lambda_{\rm sun}$ (see
Table~\ref{tab:chiq}). We find that $\chi^2$ is typically of the order
of unity, and that the range in $\chi^2$ values is the same as that
obtained in the data-models comparison. Small values of $\chi^2$ are
found for the comparison of model $q=0.8$ with $q=0.9$ and with
$q=1.25$, as well as for the comparison of $q=1.00$ with
$q=1.11$. This is most likely due to the fact that the sky projected
orbital paths of the center of mass of Sgr overlap almost perfectly up
to $\Lambda_{\rm sun} \sim 100^{\rm o}$ for each of these two groups of
models.

\begin{table}
\begin{center}
\begin{tabular}{ccccccc}
\hline
     & 0.90 & 1.00 & 1.11 & 1.25 & DATA   \\
\hline
0.80 & 0.177 & 0.706 & 0.679 & 0.276 & {\bf 0.747 (0.401)} \\
0.90 &       & 0.938 & 0.678 & 0.269 & {\bf 1.034 (0.726)} \\
1.00 &       &       & 0.061 & 1.461 & {\bf 0.937 (0.797)} \\
1.11 &       &       &       & 1.101 & {\bf 0.979 (0.806)} \\
1.25 &       &	     &       &       & {\bf 1.410 (0.621)} \\
\hline
\end{tabular}
\caption{Reduced $\chi^2$ for the different experiments. The first 4
columns are obtained by comparing the different models to one
another. The last column corresponds to the data to model
$\chi^2_{red}$. The value between brackets is obtained when the bin at
$\Lambda_{\rm sun} = 52.5^{\rm o}$ is not considered.}
\label{tab:chiq}
\end{center}
\end{table}


The extent of the stellar stream observed by M04 can be compared to
our simulations to derive its age, i.e. the time elapsed since the
stars became unbound from the Sgr dwarf. Since mass loss
preferentially occurs during pericentric passages, the age of a
(portion of a) stream may also be measured by the number of
pericentric passages that have passed since its formation. This
quantity can easily be derived for the streams in our simulations.
The material found in the range $\Lambda_{\rm sun} = [20^{\rm o},
150^{\rm o}]$ has been lost in the last 3 passages: $\sim 0.1$ Gyr ago
(dominates close to the bound core), $\sim 0.85$ Gyr ago and $\sim
1.6$ Gyr ago (dominant at $\Lambda_{\rm sun} \gsim 100^{\rm o}$).
Uncertainties in this estimate can be attributed to incomplete
knowledge of the orbit of the dwarf or of its mass-loss rate. The
former, however, is well constrained by the available data (its
average radial period lies in the range 0.7 -- 0.76 in our
models). The mass-loss rates are, on the other hand, constrained by
the velocity dispersion found in the streams. Higher mass-loss rates
than those found in these simulations would lead to thicker/hotter
streams and at a given location, the stream will be younger. On the
other hand, lower mass-loss rates would lead to colder structures
(since the particles lost would have a smaller range of energies)
which would be older on average. Inspection of Fig.~\ref{fig:vr}
suggests that the uncertainty in the quoted values cannot be larger
than $\pm 1$ pericentric passage, that is $\pm 0.75$ Gyr.

\subsection{Older streams}

\begin{figure}
\begin{center}
\includegraphics[height=0.27\textheight]{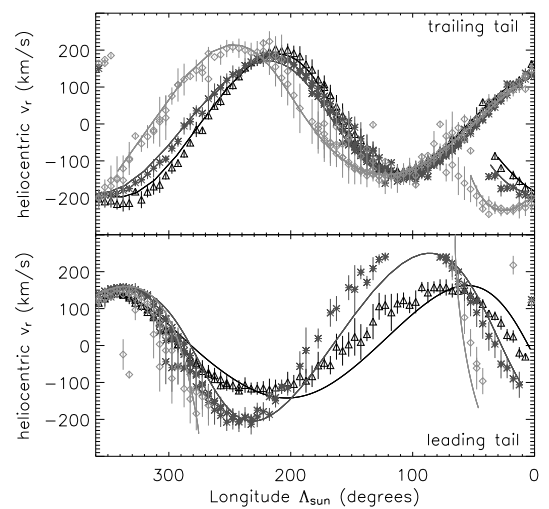}
\end{center}
\caption{The predicted mean heliocentric radial velocity measured in
bins of $\Delta \Lambda_{\rm sun} = 5^{\rm o}$ width, for particles lost in
the last 4 Gyr for models with $q=0.8$ (light gray, diamonds),
$q=1.00$ (dark gray, asterisks) and $q=1.25$ (black, triangles). The
solid curves indicate the actual trajectory followed by the center of
mass in the simulations. Note that while the overall behavior of the
latter is similar, the actual mean value at each longitude bin can be
quite different. The error bars indicate the velocity dispersion
around the mean in each bin. \label{fig:older}}
\end{figure}

How old do streams have to be before they show noticeable
dissimilarities due to variation in the Galactic dark halo shape? The
kinematics of particles lost between 2 and 4 Gyr ago (gray dots in
Fig.~\ref{fig:vr}) already exhibit important and measurable
differences.  These are further highlighted in Figure~\ref{fig:older},
which plots the trend in the heliocentric radial velocity as a
function of longitude $\Lambda_{\rm sun}$ for three different cases:
$q=0.8$ (light gray; diamonds), $q=1.00$ (dark gray; asterisks) and
$q=1.25$ (black; triangles) for particles lost in the last 4 Gyr. The
error bars indicate the expected velocity dispersion around the mean
heliocentric radial velocity in a given longitude bin. The different
models are clearly distinguishable at $\Lambda_{\rm sun} \sim 300^{\rm o}$
in the trailing stream, and for $\Lambda_{\rm sun} \sim 240^{\rm o}$ in the
leading stream.


\begin{figure}[t]
\begin{center}
\includegraphics[height=0.27\textheight]{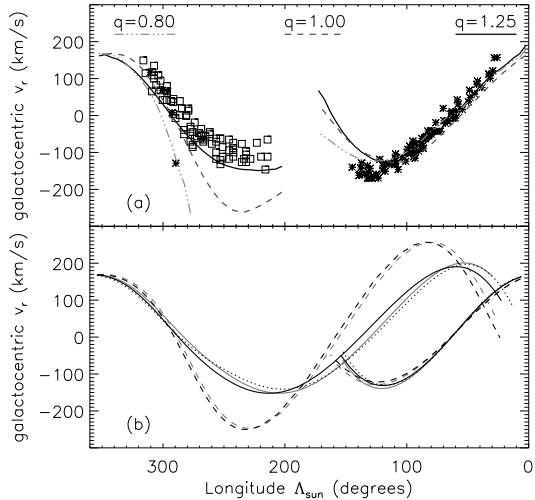}
\end{center}
\caption{{\it Panel (a)}: The predicted mean galactocentric radial
velocity vs $\Lambda_{\rm sun}$ measured in bins of $\Delta
\Lambda_{\rm sun} = 5^{\rm o}$ width, for particles lost in the last 4
Gyr of evolution for models with $q=0.8$ (light gray; dot-dashed),
$q=1.00$ (dark gray; dashed) and $q=1.25$ (black; solid). The
asterisks correspond to the data published by M04, while the open
squares are from Law et al. (2003).  {\it Panel (b)}: The
galactocentric radial velocity vs $\Lambda_{\rm sun}$ for test
particle orbits integrated in different potentials: $q = 1$ (dashed),
$q = 1.25$ (solid) and $q = 1.5$ (dotted).  The black curves
correspond to the original parameters, while the dark gray to halving
the disk mass.
\label{fig:prol}}
\end{figure}

In fact Law et al. (2003) show in their Figure 1a, the radial
velocities of a set of M giants from the 2MASS catalog located in the
leading stream at $\Lambda_{\rm sun} = [200^{\rm o}, 360^{\rm o}]$. We
have digitized this plot (since the data are not yet available in
tabular form) and show the corresponding data points as open squares
in the top panel of Figure~\ref{fig:prol}. The mean radial velocities
measured as function of $\Lambda_{\rm sun}$ for the most oblate, the
spherical and the most prolate models are overplotted for
comparison. In order to reproduce the behavior of the radial velocity
data around $\Lambda_{\rm sun} \sim 240^{\rm o}$ in the leading
stream, a prolate halo is required. A clear trend exists in the sense
that the smaller the axis ratio of the dark halo, the more negative
the mean galactocentric radial velocity at this location is, and
hence, the less likely the model becomes. If the data in Law et
al. (2003) are confirmed, this would be the {\it first direct evidence
supporting a prolate shape for the dark halo of our Galaxy}.


While the $q=1.25$ model gives the best match to this new dataset, it
seems clear that an even more elongated halo could yield a better
fit. Another option could be to reduce the mass of the disk by a small
amount while keeping this axis ratio fixed. We explore these two
possibilities using test particle orbits. The results are presented in
the bottom panel of Figure~\ref{fig:prol} for the longitudes probed by
the leading and trailing streams traced so far. The black curves
correspond to our original set of parameters: $M_{\rm disk} = 10^{11}
M_\odot$, $v_{\rm halo}= 131.5$ km/s, and $d = 12$ kpc, for the cases
$q = 1$ (dashed), $q = 1.25$ (solid) and $q = 1.5$ (dotted).  The dark
gray curves have been computed with a lower disk mass $M_{\rm disk} =
5 \times 10^{10} M_\odot$, while keeping the $v_{LSR}$ and the
circular velocity at 200 kpc fixed (this changes $v_{\rm halo}= 137$
km/s and $d = 3.5$ kpc). Therefore we conclude that reducing the mass
of the disk has a negligible effect on the orbital path. Moreover, the
very prolate $q=1.5$ halo seems to be too elongated and does not trace
well the trends in the data shown in the top panel of
Figure~\ref{fig:prol}.

\section{Summary}

There is good agreement between the models and the stars in the
trailing stream of Sgr observed by M04. The comparison confirms that
the different models can scarcely be distinguished from one another
when only the dynamics of the youngest streams ($\lesssim 1.6$ Gyr of
age) are analyzed, since all of them reproduce to similar extent the
radial velocity trends observed in the data. 

To make reliable estimates of the shape of the dark halo of the Galaxy
slightly older streams are needed. The Sgr leading stream data
published by Law et al. (2003) encompass such older debris ($\sim 2 -
4$ Gyr of dynamical age). This dataset strongly suggests that the dark
halo of the Milky Way is prolate and that the average density axis
ratio $\langle c/a \rangle_\rho$ within the orbit of Sgr is probably
larger than 5/3 ($q=1.25$) but smaller than 12/5 ($q=1.5$).

This result has a range of important implications. It rules out models
in which the dark matter is distributed as the baryons in our Galaxy,
such as for example in the form of cold $H_2$ clumps in the outer disk
(e.g. Pfenninger \& Combes, 1994). Moreover, it poses a big problem
for MOND because requires the MOND force field -- which depends on the
(oblate) distribution of luminous matter in the Galaxy -- to mimic the
gravitational effect of a prolate halo. Finally, a prolate shape could
go a long way in explaining the Holmberg effect (e.g. Sales \&
Garcia-Lambas 2004): the excess of satellites observed along the minor
axis of disk galaxies may just be reflecting the shape of the
underlying mass distribution.

\acknowledgements

I would like to thank Heather Morrison, Paul Harding, Ed Olszewski,
Simon White and Mariano Mendez for very enjoyable discussions, and the
anonymous referee for a very constructive report.

\end{document}